# A non-linear convex cost model for economic dispatch in microgrids

Vikram Bhattacharjee[1], Irfan Khan[1,a,*]

[1]Department of Electrical and Computer Engineering, Carnegie Mellon University, Pittsburgh, PA, USA

* Corresponding Author: Tel.: +412-628-5609

[a] Equal Contribution to First Author

Email Addresses : vbhattac@andrew.cmu.edu (V.Bhattacharjee);
irfank@andrew.cmu.edu (I.Khan)

# A non-linear convex cost model for economic dispatch in microgrids

Vikram Bhattacharjee[1,*], Irfan Khan[1,a,*]

[1]Department of Electrical and Computer Engineering, Carnegie Mellon University, Pittsburgh, PA, USA

**Abstract:**

This paper proposes a convex non-linear cost saving model for optimal economic dispatch in a microgrid. The model incorporates energy storage degradation cost and intermittent renewable generation. Cell degradation cost being a non-linear model, its incorporation in an objective function alters the convexity of the optimization problem and stochastic algorithms are required for its solution. This paper builds on the scope for usage of macroscopically semi-empirical models for degradation cost in economic dispatch problems and proves that these cost models derived from the existing semi-empirical capacity fade equations for LiFePO$_4$ cells are convex under some operating conditions. The proposed non-linear model was tested on two data sets of varying size which portray different trends of seasonality. The results show that the model reflects the trends of seasonality existing in the data sets and it minimizes the total fuel cost globally when compared to conventional systems of economic dispatch. The results thus indicate that the model achieves a more accurate estimate of fuel cost in the system and can be effectively utilized for cost analysis in power system applications.

*Keywords: Capacity fade, Charging/Discharging Efficiency, Fuel Savings, Economic Dispatch*

\* Corresponding Author: Tel.: +412-628-5609

[a] Equal Contribution to First Author

Email Addresses : vbhattac@andrew.cmu.edu (V.Bhattacharjee);
irfank@andrew.cmu.edu (I.Khan)

**Nomenclature**

| | | | |
|---|---|---|---|
| $P_{G-L}$ | Variable representing power flow from the diesel generator to the load at any instant (kW) | $P_{G-L}^{min}, P_{G-L}^{max}$ | Minimum and maximum bounds of power flow from the generator |
| $a, b$ | Fuel generation cost coefficients | $P_L$ | Load demand of the network (kW) |
| $P_{PV-L}$ | Variable representing power flow from the PV array to the load at any instant (kW) | $Q_C$ | Energy storage device capacity during charging (kWh) |
| $P_{PV-ES}$ | Variable representing power flow from the PV array to the energy storage device at any instant (kW) | $Q_D$ | Energy storage device capacity during discharging (kWh) |
| $P_{ES-L}$ | Variable representing power flow from the energy storage device to the load at any instant (kW) | $t_\infty$ | Life time of the energy storage device (hrs) |
| $P_{pv}$ | Variable representing power output from the PV array at any instant (kW) | $V_0$ | Terminal Voltage of the energy storage device (V) |
| $A_c$ | Variable representing area of the PV array (m²) | $R$ | Internal Resistance of the energy storage device |
| $\eta_{pv}$ | Efficiency of material of PV array | $\eta_c, \eta_d$ | Charging and discharging efficiency of the energy storage device |
| PV | Photovoltaic Generator | SOC | State of Charge |
| $SOC^{min}$ | Minimum State of Charge of the storage device | $SOC^{max}$ | Maximum State of Charge of the storage device |
| DOD | Depth of Discharge of the energy storage device | $I_{pv}$ | Hourly solar irradiation on the PV array (kWh/m²) |
| DG | Diesel Generator | BESS | Battery Energy Storage System |
| $G_1$ | Unit cost of energy consumption for generator. | $G_3$ | Unit cost of energy flow into the energy storage device |
| $G_2$ | Unit cost of energy generation from PV array. | $G_4$ | Unit cost of energy flow away from the energy storage device |

## 1. Introduction

The World Energy Outlook (WEO) database on electricity access states that till 2016, an estimated 1.2 billion population (16% of the global population) do not have access to electricity. Around 95% of the deprived population belong to the remote areas. [1]. In the present scenario, efforts are being made to design efficient microgrid solutions to aid penetration of power supply in these remote areas through components like energy storage, photovoltaic (PV) generation and diesel generators [2]. These efficient designs can be achieved through optimal sizing grid components [2], operation cost minimization [3-4], optimal energy storage sizing [5], real time energy management using stochastic techniques [6-11], quadratic approximations to the optimal power flow problem [12], deploying distributed algorithms for efficient convergence rates [13], designing load management strategies [14,15] and analyzing techno-economic feasibility [16]. Of all these available resources, the techno-economic feasibility model has been widely explored for achieving viable off grid solutions [17]. Wies et al. [18] developed a techno-economic feasibility model of hybrid system which reduced the operation cost of the system drastically. Techno-economic feasibility is judged by the economic dispatch of the diesel generators in the grid under operational constraints.

Economic dispatch can be defined as a process to retrieve generation in a grid at minimum cost [19]. With the incorporation of more and more system constraints, energy management at reduced cost becomes a complicated problem to solve. The present day economic dispatch problems consider incorporation of system constraints like security [20], appropriate power levels [21], daily averages of solar irradiation [22] and operational cost of the energy storage devices in the hybrid microgrid system [23]. Renewable generation cost can be correlated to linear functions of output power from their respective devices [24] and they contribute to the load demand of the system whose stochastic nature can be forecasted using sophisticated techniques such as neural networks [25], ensemble learning [26] and reinforcement learning models [27]. However, the operational cost of the energy storage devices heavily depends on their respective capacities that fade over time due to varying charge and discharge efficiencies during subsequent charge/discharge cycles [28,29]. Capacity degradation is thus one such inherent property which incorporates complexity in designing hybrid systems. The most accurate models for battery degradation have been modelled using electrochemical models as PDE observers [30,31]. However, the states of the system heavily rely on the stability of the estimator. The other methods include designing non-linear empirical models to relate high level system parameters like state of charge and voltage based on experimental observations [32]. These models are highly non-linear in nature and depend on stochastic algorithms for solution [33].

Present research considers economic dispatch under the influence of non-uniform load distribution and non-linear battery degradation models. Incorporation of these variations lead to the development of non-linear optimization problems which require genetic algorithms [21] and inbuilt commercial software packages for solution [34, 35]. A commercial program (Hybrid Optimization by Genetic Algorithm-HOGA) [34] determines the optimal configuration of a hybrid PV-DG system using genetic algorithms. The package incorporates non-linear characteristics of system components like load demand and uncertainty in renewable energy supply. A commercial software called HOMER (Hybrid Optimization Model for Electric Renewables), developed by National Renewable Energy Laboratory, USA is used to judge the dimensioning of hybrid power systems based on system cost, operation constraints and load demand through hourly simulations [35]. In [36], battery degradation cost was considered in designing a hybrid standalone system which aims to minimize the total operating cost. A genetic algorithm was utilized to solve the problem and the strategy achieved a net cost reduction of 37.7% when compared to generic load following strategies [37]. Although these studies are extensive, they are strictly dependent on stochastic methods for solution. The non-linearity of these characteristics impose uncertainty on judging whether the costs are globally optimal or not. Therefore, it is techno-economically more feasible to design convex optimization problems for these hybrid systems in order to get an accurate estimate of the total system cost and escalate fuel savings. Tazvinga et al. [38] considered dynamic variation of load demand to calculate a dispatch strategy to minimize the operation cost. They formulated the techno-economic feasibility problem as a convex problem and showed that dynamic variation of fuel costs drastically affects cost savings of the generator. However, their model did not consider the cost of energy storage and intermittent renewable generation which alters the optimality conditions of the design problem.

From the above literature, it is evident that the economic dispatch model is a non-linear problem for optimization and incorporation of parameter variation may result in sub-optimal solutions. The paper develops an optimal pathway towards finding the minimum cost solution for such dynamic non-linear fuel saving based cost models. A non-linear operational convex cost model incorporating energy storage dynamics and intermittent renewable generation has been proposed in this paper. Cell dynamics has been incorporated using a well-established semi-empirical relation between cell capacity and its rated power for a class of an energy storage device. The paper further proves the convexity of the non-linear cost model and provides the conditions under which the conditions of convexity will hold.

## 2. Contribution to Research

The phenomenon of undercharge and under-discharge reduces the available capacity of the battery pack when compared to its nominal value [39]. This reduction in capacity affects the operation of energy management problems where BESS forms an integral part. In order to address an efficient dispatch strategy, researchers integrate battery degradation cost in their objective for optimization. The degradation cost modeling is an important avenue for research in this scenario. There are mainly two approaches to modeling degradation cost. The first being the incorporation of a decaying degradation trend through estimation of life cycle curves (also known as DOD Swing) and the second being the incorporation of cell degradation cost as a stochastic process. Xiao et al [40] utilized a non-linear decaying lifetime cumulative degradation cost model based current DOD status for a real time DC microgrid scheduling. In [41] an optimal utilization strategy was designed for minimizing operational cost through additional degradation cost modeling by the same technique as [40]. They solved the optimization problem using both GAMS and CPLEX solvers. In [42] a similar life cycle estimation technique was utilized for combining storage systems with wind generation units. They utilized the CPLEX framework for the solution. However, the solvers were reported to exhibit a negligible but finite optimality gap. The other technique refers to introduction of cell degradation as a stochastic process. In [43], the battery pack degradation was modeled as a Weiner process.

However, the technique has not been applied to economic dispatch models in the past. Both the techniques described in the literature are stochastic in nature and are theoretically poised to provide sub-optimal solutions. Since the sources of sub-optimality in degradation cost based economic dispatch problems lie in the non-linear models life cycle estimation models, the goal of the paper is to answer three key questions.

**Q.1** Can there exist optimal solutions to such economic dispatch problems which encourage energy storage with zero optimality gap?

**Q.2** If they exist, what are the conditions for them to exist?

**Q.3** How do we solve those optimization problems?

Based on the above questions, the three main contributions of the paper are as follows: -

**2.1 A convex semi empirical degradation cost model for economic dispatch**

A recent paper on the application of degradation maps on power system as a benchmark reference, discusses the implementation of semi-empirical cost models for implementation in power system frameworks [44]. They use a system identification technique to map the lost charge capacity to the state of charge of the system. With simultaneous understanding the patterns of battery usage from life cycle curves and cycle test data, the final step is a convexification process which produces convex degradation maps that are readily implementable in power system architecture. This paper further builds on the scope for usage of a macroscopically semi-empirical models for degradation cost in economic dispatch. The literature consists of semi-empirical models for available capacity for $LiFePO_4$ battery cells during both cases of charging and discharging [39]. The paper derives the degradation cost models from the existing semi-empirical equations and proposes the fact that the degradation cost models for $LiFePO_4$ battery cells are convex under certain operating conditions where they do not require the previously explained preprocessing steps. We apply the derived models on an economic dispatch strategy to calculate the global solution and cost savings.

**2.2 Description of the solution objective**

We use the cost model described in [45] to calculate our cost savings. We incorporate degradation cost into the objective function by deriving the relations for the dynamic charging and discharge efficiencies according to the semi-empirical models for available capacity.

## 2.3 Solution to these non-linear optimization problems

The fully distributed Alternating Directional Method of Multipliers or simply the ADMM technique [46] has been used to solve the non-linear convex optimization problem. The ADMM technique solves an optimization problem by decomposing it down to smaller sub problems. It is a procedure which enhances the coordination between local sub problems to achieve the global solution to the large master problem under consideration. Furthermore, the optimal cost solution has been tested on two test cases for available load demand data. The first test case consists of a data set for a winter and summer weekday. The second test case consists of the electrical consumption data for the whole year taken at an interval of 15 minutes. The cost reduction has been compared with two different strategies:-

a) Diesel Based System: A system which has no other component except the diesel generator which meets the entire load demand. This basically means that the optimization problem is an economic dispatch with power generation limit constraints and supply demand balance constraints only.

b) Hybrid System without Pack Degradation: A system which has the same objective as the given problem but the effect of battery pack degradation is not included in the cost for energy storage or in other words the charging and discharging efficiencies are not dynamic. This reduces the objective to a quadratic problem which can be solved using the "quadprog" or CVX routine in MATLAB.

The salient features of the proposed formulation have been described below:-

1. The cost model been designed keeping in mind the load demand and generation balance.
2. The model incorporates the effects of battery pack degradation and intermittent renewable generation.
3. The non-linear model has been proved to be convex under specific operating conditions in order to ensure that a global optimal solution is retrieved.

The rest of the paper has been organized as follows. Chapter 3 discusses the components of the hybrid model and the dynamics of each component of a hybrid model. Chapter 4 discusses the non-linear cost model and the conditions of convexity. Chapter 5 discusses the algorithm for solving the optimization problem. Chapter 6 discusses the experimental results while Chapter 7 and 8 highlights the key limitations and conclusions of the paper respectively.

**3. Model Definitions**

In this paper, matrices are represented by bold and italic letters, and scalar quantities are neither bold nor italic. Functions and variables are shown as italic letters (not bold).

**3.1 Hybrid System**

A hybrid system has been considered to be made up of three main components, the photovoltaic (PV) system, the Diesel Generator (DG) and the battery energy storage system (BESS). In the system the load demand is met by adaptive coordination of the above three components. The BESS acts as the backup storage option which provides support when the PV output is not enough to meet the load demand. The best feature of a hybrid system is that the BESS stores the excess energy from the PV, thereby facilitating maximum energy utilization. The scheduling of the DG occurs when the energy flow from the PV and the BESS is not enough to meet the load demand. The techno-economic feasibility behind the working condition of a hybrid system is maximized by minimizing the excess usage of the DG. The proposed simulation process in terms of the input, output and simulation model along with constraints included is shown in Fig.1. We explain the formulation of the components used in this paper in the next following sections.

In this paper the power flow in the hybrid system has been defined according to the Fig 1. $P_{G-L}$ and $P_{PV-L}$ represent the energy flow from DG and the PV array to the load while $P_{PV-ES}$ and $P_{ES-L}$ represent the energy flow from the PV to the battery and from the battery bank to the load respectively. For clarity purposes, the sign of energy flow both to and from the battery bank has been considered to be positive. We further consider a lithium-ion battery model for the BESS system keeping in mind its wide scope for applications in the power and energy sectors.

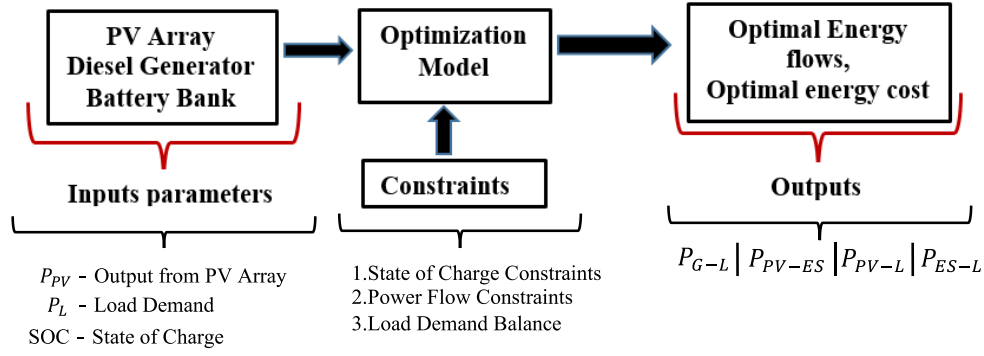

Fig 1. Hybrid System

### 3.2 PV Array model

The power output ($P_{pv}$) from the PV array model is directly proportional to the area of the array ($A_c$), the generator efficiency ($\eta_{pv}$) and the hourly incident solar irradiation ($I_{pv}$). The relation has been shown in Eq.(1).

$$P_{PV} = A_c \eta_{PV} I_{PV} \tag{1}$$

Furthermore, the energy flow from the PV generator ($P_{pv}$) is less than or equal to the total sum of the energy flow from the PV to the load and the energy flow from PV into the BESS. The relation is shown in Eq.(2).

$$P_{PV-L}(t) + P_{PV-ES}(t) \leq P_{PV}(t) \tag{2}$$

### 3.3 DG model

The DGs provide back-up power supply when the available energy from the PV and the BESS is not enough to meet the load demand. Considering a line loss of 5%, the load demand is equal to the energy flow from the DG, battery and the PV array. For a single network, the overall energy flow from DG, battery and the renewable generation will be equal to the overall load demand of the system. The relation for the global energy balance has been shown in Eq.(3).

$$P_{G-L}(t) + P_{PV-L}(t) + P_{ES-L}(t) = 1.05 P_L(t) \tag{3}$$

In this paper, a 5 kVA DG unit of the hybrid system has been specified to operate between lower and upper limits $P_{G-L}^{min}$ and $P_{G-L}^{max}$ respectively. The constraint has been shown in Eq. (4).

$$P_{G-L}^{\min}(t) \leq P_{G-L}(t) \leq P_{G-L}^{\max}(t) \tag{4}$$

Generally, the prime objective of a hybrid system is to minimize the intermittent usage of the DG in order to meet the load demand of the system. The load demand is expressed as a function of seasonal and diurnal variation.

### 3.4 Battery Bank

The battery bank provides the power output when the output of the PV system is not enough to meet the load demand. During subsequent charge/discharge cycles the capacity of the cells of the battery bank falls below its nominal value due to effects of undercharge, under-discharge and internal losses and side reactions [39]. The available capacity at any instant for each cell during both charging and discharging processes is thus expressed as a fraction of the total rated capacity of the cell. In [39], the available capacities for both the cases have been expressed as a product of the nominal capacity of the cell along with empirical functions of power flow from the energy storage device. The relations for the available capacity for the charging and discharging cases are given by Eq.(5) and (6) respectively. If $Q_o$ be the total initial capacity of the pack, Eq.(5) shows that the available capacity ($Q_c(P_{PV\text{-}ES})$) is related to an empirical quadratic function of power input $P_{PV-ES}$ to the storage device. The function is dependent on coefficients u and v which are constants specific to the device. On the other hand, Eq. 6 shows a similar relation between the available capacity during the discharge process ($Q_d(P_{ES\text{-}L})$) and the power output $P_{ES-L}$ from the energy storage device. The function in the second case also depends on the maximum rated output power flow from the storage device ($P_{max}$).

$$Q_c(P_{PV-ES}) = (1 - uP_{PV-ES} - vP_{PV-ES}^2)Q_0 \quad (5)$$

$$Q_d(P_{ES-L}) = \tanh\left(\frac{P_{max} - P_{ES-L}}{\sqrt{P_{max} + P_{ES-L}}}\right) Q_o \quad (6)$$

The life of the battery is defined as the time taken by the cell to charge it to its full capacity when charged at a constant current *I*. It is denoted as $t_\infty$ which is expressed as a fraction of the total capacity $Q_o$ and the discharge current *I* according to Eq. (7).

$$t_\infty = \frac{Q_o}{I} \quad (7)$$

The charge current $I$ flowing through the equivalent circuit of the cell is related to the open circuit voltage $V_0$ and the internal resistance $R$ by Eq. (8).

$$I = -\frac{V_0}{2R} \pm \sqrt{\left(\frac{V_0}{2R}\right)^2 + \frac{P_{PV-ES}}{R}} \tag{8}$$

When the total capacity is less than the nominal capacity due to effects of undercharge, the life of the pack is expressed as a function of the available capacity, $Q_c(P_{PV\text{-}ES})$, in place of the nominal capacity $Q_o$ when charged at a constant rate $P_{PV-ES}$. Thus for the charging process, the battery life time will be expressed according to the relation in Eq.(9).

$$t_\infty = \frac{Q_c(P_{PV-ES})}{I} \tag{9}$$

The total energy stored in the battery due to charging at a constant rate $P_{PV-ES}$ at infinite time for a 900mAh Tenergy lithium-ion is expressed in Eq.(10) [39].

$$E_c(P_{PV-ES}) = Q_0 V_0 + \left(\frac{V_0}{2R} - \sqrt{\left(\frac{V_0}{2R}\right)^2 + \frac{P_{PV-ES}}{R}}\right) Q_c(P_{PV-ES}) R \tag{10}$$

The charging efficiency ($\eta_c$) is defined as the ratio between the energy stored in the battery at infinite time to the rated capacity ($Q_0 V_0$). Hence $\alpha = \frac{2R}{V_0^2}$ is defined as a constant where the parameter $2\alpha P_{PV-ES}$ is used to generate the operational Ragone curves [28] for the energy storage device under consideration. Substituting Eq.(5) in Eq.(10), we get Eq.(11).

$$\eta_c(P_{PV-ES}) = 1 + \frac{1}{2}(1 - \sqrt{1 + 2\alpha P_{PV-ES}})(1 - u P_{PV-ES} - v P_{PV-ES}^2) \tag{11}$$

The life of the battery is defined as the time taken by the cell to discharge completely at a constant current $I$. When the total capacity is less than the nominal capacity due to effects of effects of underdischarge, the life of the pack is expressed as a function of the available capacity, $Q_d(P_{ES\text{-}L})$, in place of the nominal capacity $Q_o$ when charged at a constant rate $P_{ES-L}$. Thus for the discharging process, the battery life time will be expressed as in Eq.(12).

$$t_\infty = \frac{Q_c(P_{PV-ES})}{I} \tag{12}$$

The discharge current $I$ is related to the open circuit voltage $V_0$ and the internal resistance $R$ by Eq.(13).

$$I = \frac{V_0}{2R} - \sqrt{\left(\frac{V_0}{2R}\right)^2 - \frac{P_{ES-L}}{R}} \tag{13}$$

Similarly, the discharge energy available for the load at infinite time is defined as the ratio between the energy discharged from the load till infinite time to the rated capacity ($Q_0 V_0$). The discharging efficiency ($\eta_d$) is defined as the ratio between the energy discharged from the battery at infinite time to the rated capacity ($Q_0 V_0$). The relation has been shown in Eq.(14) [28].

$$E_d(P_{ES-L}) = \frac{2R Q_d(P_{ES-L}) P_{ES-L}}{V_0 - \sqrt{V_0^2 - 4R P_{ES-L}}} \tag{14}$$

Substituting Eq.(2) in Eq.(14), we get Eq.(15).

$$\eta_d(P_{ES-L}) = \frac{\frac{2R}{V_0^2} \tanh\left(\frac{P_{max} - P_{ES-L}}{\sqrt{P_{max} + P_{ES-L}}}\right) P_{ES-L}}{1 - \sqrt{1 - \frac{4R P_{ES-L}}{V_0^2}}} \tag{15}$$

Now since the internal resistance is low, we can approximate $\sqrt{1 - \frac{4R P_{ES-L}}{V_0^2}} \approx 1 - \frac{2R P_{ES-L}}{V_0^2}$ by neglecting the terms of higher indices. This approximation holds true until the higher order terms affect the solution. Mathematically, the solution methodology will remain unaffected till the order of internal resistance raises to the order of $10^{-2}$. Therefore the discharge efficiency can be expressed as in Eq.(16).

$$\eta_d(P_{ES-L}) = \tanh\left(\frac{P_{max} - P_{ES-L}}{\sqrt{P_{max} + P_{ES-L}}}\right) \tag{16}$$

At a given $t$-th hour, the state of charge (SOC) of the battery bank is dependent on the state of charge at the previous hour and the net energy flow during subsequent charge and discharge. The SOC at time $t$ given by Eq.(17) transforms to Eq.(18), where SOC(0) is the initial SOC of the battery, $\sum_{\tau=1}^{t} \eta_c P_{PV-ES}(\tau)$ is the energy flow into the battery at time *t*, and $\sum_{\tau=1}^{t}\eta_d^{-1}P_{ES-L}(\tau)$ is the energy flow away from the battery at time *t*. The lowest operational limit for the SOC of any pack in the system is given by $SOC^{min}$ and upper limit of the state of charge in the system is given by $SOC^{max}$ (Eq.(19)). The lower limit of the SOC is related to the upper limit by the depth of discharge (DOD) of the battery pack. The Depth of Discharge (DOD) is the maximum allowable capacity that can be discharged from a fully charged cell. Below the allowable capacity, we consider the cell to be out of service. The DOD concept has been used to calculate the lower operational limit for the SOC in the system. The relation has been shown in Eq.(20).

$$SOC(t) = SOC(t-1) + \eta_c P_{PV-ES}(t) - \eta^{-1}{}_d P_{ES-L}(t) \qquad (17)$$

$$SOC(t) = SOC(0) + \sum_{\tau=1}^{t}\eta_c P_{PV-ES}(\tau) - \sum_{\tau=1}^{t}\eta^{-1}{}_d P_{ES-L}(\tau) \qquad (18)$$

$$SOC^{min} \leq SOC(t) \leq SOC^{max} \qquad (19)$$

$$SOC^{min} = (1-DOD)SOC^{max} \qquad (20)$$

In this paper an additional constraint has been introduced in order to ensure there is no simultaneous charging and discharging of the BESS at the same instant. A weighted inequality consisting of the power flow components during charge and discharge conditions has been presented in Eq.(21). The inequality consists of two coefficients $n_1$ and $n_2$ ($n_1, n_2 \in I^+$) which vary with every time step and $h^{max}$ represents a very small positive number.

$$n_1 P_{PV-ES}(t) + n_2 P_{ES-L}(t) < h^{max} \qquad (21)$$

The coefficients are chosen in such a manner that when the magnitude of the charging component is higher than that of the discharge component, $n_2$ becomes unity and $n_1$ becomes zero. This approach depresses the presence of the discharge component when both the components are positive. Mathematically, assuming $P_{PV-ES} > P_{ES-L} > 0$, the equation Eq.(21) can be rewritten as Eq.(22) which provides the outcome of the regime switching approach.

$$P_{ES-L}(t) < h^{max} \Rightarrow P_{ES-L}(t) \approx 0 \tag{22}$$

The value $h^{max}$ is a very small positive number which thereby establishes that $P_{ES-L}$ is suppressed in the system when $P_{PV-ES} \neq 0$.

## 4. Optimization Model

The objective of this paper is to reduce the fuel cost of the generator. The fuel cost of the generator is compensated by the cost of energy storage and renewable generation. The cost of the operation of the DGs is a quadratic function of the active power flow for meeting the load demand [47] and it is given by Eq. (25).

$$J_1 = G_1(aP_{G-L}^2(t) + bP_{G-L}(t)) \tag{25}$$

The cost saved due to power generation from a PV array in time $t$ is given by Eq. (26).

$$J_3 = G_2 P_{PV-L}(t) \tag{26}$$

If $\eta_c P_{PV-ES}(t)$ be the actual energy flow into the battery and $\eta_D^{-1} P_{ES-L}(t)$ is the energy flow away from the battery at time $t$, then the cost model for the BESS component will be given by Eq.(27).

$$J_2 = G_3 \eta_c P_{PV-ES}(t) - G_4 \eta_d^{-1} P_{ES-L}(t) \tag{27}$$

The overall function depicting cost of operation of the generator of the microgrid is given by Eq.(28)

$$Cost = \sum_{t=1}^{N} w_1 J_1 - w_2 J_2 - w_3 J_3 \tag{28}$$

The coefficients $G_1, G_2, G_3$ and $G_4$ are the unit costs of energy flow expressed in US\$/kW. $w_1$, $w_2$ and $w_3$ are the weights for the corresponding hybrid system components.

The reference optimization scheme further constricts the states to being positive and having an upper and a lower bound. The formulation along with the equality and inequality constraints denoted by Eqs. (2), (3), (4) and (21) is given by (27). The proposed scheme in (29) is globally optimal (*for proof see Section 4*).

$$\min \sum_{t=1}^{N} w_1 J_1 - w_2 J_2 - w_3 J_3$$

$$\text{s.t.} \quad \begin{aligned} & P_{PV-L}(t) + P_{PV-ES}(t) \leq P_{PV}(t) \\ & P_{G-L}(t) + P_{PV-L}(t) + P_{ES-L}(t) = 1.05 P_L(t) \\ & n_1 P_{PV-ES}(t) + n_2 P_{ES-L}(t) < h^{\max} \\ & 0 \leq P_{G-L}(t) \leq P_{G-L}^{\max} \\ & 0 \leq P_{PV-L}(t) \leq P_{PV-L}^{\max} \\ & 0 \leq P_{PV-ES}(t) \leq P_{PV-ES}^{\max} \\ & 0 \leq P_{ES-L}(t) \leq P_{ES-L}^{\max} \end{aligned} \quad (29)$$

The constraint in Eq.(19) has been considered as a projection on (29). The state of charge is calculated and the projection is revised at the end of every iteration . The optimization continues with the storage component till the SOC of the battery pack remains within the specified limit. In any other case the contribution from the energy storage is considered to be zero. The relation of the projection has been shown in Eq.(30) and Eq.(31) respectively.

$$P_{PV-ES}(t) = \begin{cases} P_{PV-ES}(t), & \text{SOC}_{\min} \leq SOC(t) \leq \text{SOC}_{\max} \\ 0, & otherwise \end{cases} \quad (30)$$

$$P_{ES-L}(t) = \begin{cases} P_{ES-L}(t), & \text{SOC}_{\min} \leq SOC(t) \leq \text{SOC}_{\max} \\ 0, & otherwise \end{cases} \quad (31)$$

### 4.1 Proof of Convexity of the Optimization Problem

The problem poised in (29) is non-linear in nature due to the presence of the capacity fade terms in cost model for the energy storage component of the hybrid system. In this section, we will be proving the convexity of the non-linear cost model and will be providing the conditions under which the convexity will be preserved.

### 4.2 Convexity of the Energy Storage Cost Model

For the energy storage cost model, we have discharging and charging cost components and each are proved separately.

### 4.2.1 Convexity of the Discharge Cost Component

Since we consider the absolute values of the parameters in the problem, all the power flow terms are positive. Hence the mathematical relations in Eq(32)-(34) will hold true for positive values of $P$ and $P_{max}$.

$$0 < \tanh\left(\frac{P_{max} - P}{\sqrt{P_{max} + P}}\right) < 1 \tag{32}$$

$$\operatorname{sech}^2\left(\frac{P_{max} - P}{\sqrt{P_{max} + P}}\right) > 0 \tag{33}$$

$$\left(2P^3 + 12P_{max}P^2 + 18P_{max}^2 P\right) > 0 \tag{34}$$

A function $g(x)$ is strictly convex in its domain if and only if $g''(x) > 0 \ \forall x$. The function for cost due to discharge is given by Eq.(35). Differentiating twice with respect to the power flow terms, we get Eq.(36).

$$g(P) = P\eta_d(P)^{-1} = P\tanh\left(\frac{P_{max} - P}{\sqrt{P_{max} + P}}\right)^{-1} \tag{35}$$

$$g''(P) = \frac{\operatorname{sech}^2\left(\frac{P_{max} - P}{\sqrt{P_{max} + P}}\right)(x + y + z)}{4(P_{max} + P)^3 \tanh^2\left(\frac{P_{max} - P}{\sqrt{P_{max} + P}}\right)} \tag{36}$$

The terms $x$, $y$ and $z$ are given by Eqs.(37),(38) and (39) respectively.

$$x = \left(2P^3 + 12P_{max}P^2 + 18P_{max}^2 P\right)\tanh^2\left(\frac{P_{max} - P}{\sqrt{P_{max} + P}}\right) \geq 0 \tag{37}$$

$$y = \tanh\left(\frac{P_{max} - P}{\sqrt{(P_{max} + P)}}\right) \cdot \sqrt{P_{max} + P} \cdot \left(12P^2_{max} + 9P_{max}P + 3P^2\right) \geq 0 \tag{38}$$

$$z = \left(2P^3 + 12P_{max}P^2 + 18P_{max}^2 P\right)\operatorname{sech}^2\left(\frac{P_{max} - P}{\sqrt{P_{max} + P}}\right) \geq 0 \tag{39}$$

Adding Eq. (37), (38) and (39), we get,

$$x + y + z \geq 0 \tag{40}$$

which gives,

$$g''(P) = \frac{\text{sech}^2\left(\frac{P_{max} - P}{\sqrt{P_{max} + P}}\right)(x+y+z)}{4(P_{max} + P)^3 \tanh^2\left(\frac{P_{max} - P}{\sqrt{P_{max} + P}}\right)} \geq 0 \forall P \in [0, P_{max}] \quad (41)$$

According to the condition in Eq.(41), the discharging cost component of the non-linear objective function is convex in nature.

### 4.2.2 Concavity of the Charging Cost Component

The function for charging cost is given by Eq.(42).

$$P\eta_c(P) = P + \frac{P}{2}\left(1 - \sqrt{1 + \frac{2R_0}{V_0^2}P}\right)(1 - uP - vP^2) \quad (42)$$

Utilizing Taylor's Expansion of the rational terms in Eq.(42), we rewrite it as Eq.(43)

$$P\eta_c(P) = P - \frac{1}{2}\left(\sum_{i=1}^{\infty} \frac{(\alpha^i P^{i+1} - u\alpha^i P^{i+2} - v\alpha^i P^{i+3})}{i!}\right) \quad (43)$$

Eq.(43) is a non-linear equation and since the power flow has been considered here to be in kW, the coefficients of the higher powers of $P$ should be small enough for them to be neglected. Hence, the relaxation of the terms for preservation of concavity of the function in Eq.(43) will depend on the choice the index of $\alpha$. Considering the individual indices of the constants $u$ and $v$ and the general terms of the expression, if $z$ and $i$ be the indices of $\alpha$ and $P$ in Eq.(43), the index $i_{opt}$ for which the terms can be deemed infeasible for consideration, will be given according to the relation in Eq.(44).

$$i_{opt} = \max\{((3-z)i+3), ((3-z)i+4), ((3-z)i+6)\} \quad (44)$$

For the function in Eq.(43) to be quadratic, we will calculate the index of $z$ considering $i$ to be equal to 1. The concavity will hold if the largest index of the three components in $i_{opt}$ from Eq. (44), is less than zero, which in turn implies that $z > 9$. In that case, correlating with the theory of Ragone Plots [27], the corresponding parameter $\alpha P$ will have an index equal to -6. Thus, all the indices from Eq. (44) will be negative till the Ragone parameter $\alpha P$ increases to an order of $10^{-6}$. Hence, it can be proposed that the concavity of the charging cost model will be preserved till the Ragone parameter $\alpha P$ increases to an order of $10^{-6}$. Under the following conditions the charging cost function will be of the form as shown in Eq.(45).

$$P\eta_c(P) = P - \frac{1}{2}\alpha P^2 \qquad (45)$$

The models for the individual generator fuel cost and the cost of renewable generation are quadratic and linear in nature. Hence the individual components of the cost model in Eq.(28) have been proved to be convex functions which implies that the minimization problem under linear constraints is convex in nature. The solution to optimization problem is discussed in the following section.

## 5. Solution to the Optimization Problem

The optimization problem in Eq.(29) takes the form of Eq.(46) where $A_1, B_1, C_1$ and $D_1$ denote the coefficient matrices for the inequality constraints. The matrices $A_2, B_2, C_2$ and $D_2$ denotes the coefficient matrices for the corresponding equality constraints.

$$\begin{aligned}
\min\ & f_1(P_{G-L}) + f_2(P_{PV-L}) + f_3(P_{PV-ES}) + f_4(P_{ES-L}) \\
\text{s.t.}\ & A_1 P_{G-L} + B_1 P_{PV-L} + C_1 P_{PV-ES} + D_1 P_{ES-L} \leq c_1 \\
& A_2 P_{G-L} + B_2 P_{PV-L} + C_2 P_{PV-ES} + D_2 P_{ES-L} = c_2
\end{aligned} \qquad (46)$$

The components of the objective function are explicit from each other and convex in nature while the constraints are linear. Thus, the problem can be solved in a completely distributed manner using ADMM [46]. The inequality constraints of the optimization problem can be further converted into equality constraints through the introduction of slack variables $\varepsilon_i$ on the objective function. Hence the optimization problem transforms to a form as shown in (47) where $A, B, C, D$ and $E_i$ represent the matrices for the equality constraints.

$$\begin{aligned}
\min\ & f_1(P_{G-L}) + f_2(P_{PV-L}) + f_3(P_{PV-ES}) + f_4(P_{ES-L}) \\
\text{s.t.}\ & AP_{G-L} + BP_{PV-L} + CP_{PV-ES} + DP_{ES-L} + \sum_i E_i \varepsilon_i = c
\end{aligned} \qquad (47)$$

The corresponding augmented Lagrangian is given by (48).

$$L = f_1(P_{G-L}) + f_2(P_{PV-L}) + f_3(P_{PV-ES}) + f_4(P_{ES-L})$$
$$+ \mu^T \left( AP_{G-L} + BP_{PV-L} + CP_{PV-ES} + DP_{ES-L} + \sum_i E_i \varepsilon_i - c \right) \quad (48)$$
$$+ \frac{\rho}{2} \left\| AP_{G-L} + BP_{PV-L} + CP_{PV-ES} + DP_{ES-L} + \sum_i E_i \varepsilon_i - c \right\|_2^2$$

Using ADMM, the update process is given by Eq, (49)-(54) where the states are updated consecutively before the Lagrange multipliers are updated at each step.

$$P_{G-L}^{k+1} = \arg\min_{P_{G-L}} L\left(P_{PV-L}^k, P_{PV-ES}^k, P_{ES-L}^k, \varepsilon_i^k, \mu^k\right) \quad (49)$$

$$P_{PV-L}^{k+1} = \arg\min_{P_{PV-L}} L\left(P_{G-L}^{k+1}, P_{PV-ES}^k, P_{ES-L}^k, \varepsilon_i^k, \mu^k\right) \quad (50)$$

$$P_{PV-ES}^{k+1} = \arg\min_{P_{PV-ES}} L\left(P_{G-L}^{k+1}, P_{PV-L}^{k+1}, P_{ES-L}^k, \varepsilon_i^k, \mu^k\right) \quad (51)$$

$$P_{ES-L}^{k+1} = \arg\min_{P_{ES-L}} L\left(P_{G-L}^{k+1}, P_{PV-L}^{k+1}, P_{PV-ES}^{k+1}, \varepsilon_i^k, \mu^k\right) \quad (52)$$

$$\varepsilon_i^{k+1} = \arg\min_{\varepsilon_i} L\left(P_{G-L}^{k+1}, P_{PV-L}^{k+1}, P_{PV-ES}^{k+1}, P_{ES-L}^{k+1}, \mu^k\right) \quad (53)$$

$$\mu^{k+1} = \mu^k + \left( AP_{G-L}^{k+1} + BP_{PV-L}^{k+1} + CP_{PV-ES}^{k+1} + DP_{ES-L}^{k+1} + \sum_i E_i \varepsilon_i^{k+1} - c \right) \quad (54)$$

The algorithm for finding the least cost solution has been shown below in the following flowchart in Fig 2. From the initial values of the states, the state of charge of the system is calculated by Eq.(18). If the SOC is within limits, then the states are updated according to the Eq.(49)-(52). At each step of the update process, the non-linear equations (51) and (52) are solved with an appropriate guess by the help of the Newton Raphson Technique. After the states are updated according to Eq.(50)-(52), the corresponding slack variables and the Lagrange Multipliers are updated according to Eq.(53)-(54) respectively. The overall update process continues till a designated error tolerance is reached.

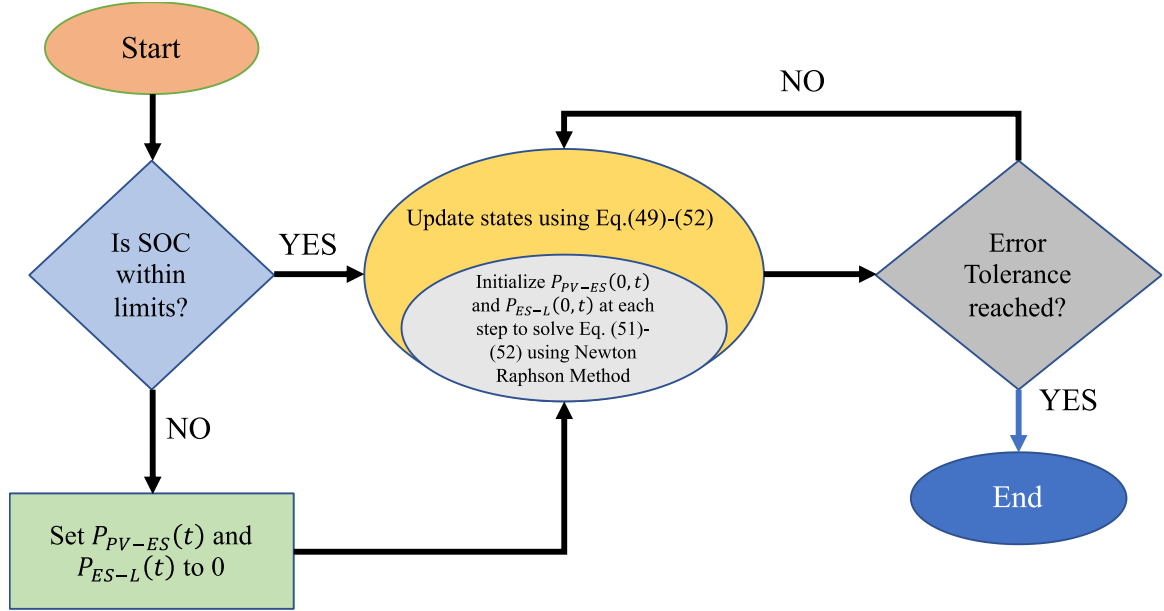

Fig.2. Algorithm Flowchart

The model parameters of the components and the values of the constants have been tabulated below in Table 1.

Table 1. Model Constants

| Parameter | Value | Unit | References |
|---|---|---|---|
| $h^{max}$ | $10^{-6}$ | - | - |
| $u$ | 0.035 | kW$^{-1}$ | [39] |
| $v$ | 0.0052 | kW$^{-2}$ | [39] |
| $P_{max}$ | 12 | kW | - |
| $w_1$, $w_2$ and $w_3$ | 1,10,0.1 | - | - |
| $G_1$,$G_2$,$G_3$ and ,$G_4$ | 1 | US\$ kW$^{-1}$ | - |
| PV Array Capacity | 4.5 | kW | - |
| Diesel Generator capacity | 5 | kVA | - |
| Depth of Discharge(*DOD*) | 50% | - | - |
| Nominal Battery Capacity | 55 | kW | [38] |
| $a$ | 0.25 | US\$/h | [38] |
| $b$ | 0.1 | US\$/kWh | [38] |

## 5.1 Test Case

In order to test the validity of the proposed formulation, we test our algorithm on two datasets.The first data set summarizes the load demand data on a hour of the day basis. The second data set features the load demand data collected over the whole year on a day of the year basis. The description of the data sets and their characteristics have been explained below in the following sections.

**5.1.1 Case 1**

The proposed formulation was tested on a load demand data set taken from rural community clinics in Zimbawe for the summer and winter weekday. The data was originally collected on the average day for 24 hours of each month and at the mid-point of every hour [38]. The data set shows that for the winter weekday, the demand is lower towards the early morning and higher towards the noon when compared to that of the summer. A schematic of the load demand data collected over 24 hours of a summer and a winter weekday has been shown below in Fig 3.

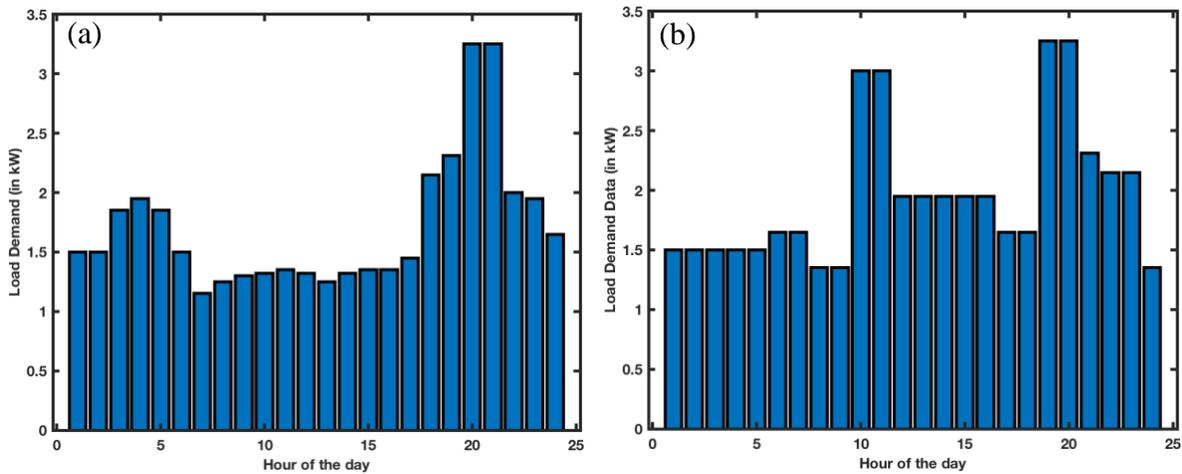

Fig.3. Load Demand Dataset for (a). Summer Weekday and (b) Winter Weekday

**5.1.2 Case II**

A larger data set for Northern Ireland was considered in the second case where the intraday 15-minute load demand data was collected for a year [48]. The data set shows annual seasonality and has been used to test if the proposed formulation captures similar trends. We calculate the average load demand for each day by taking the mean of all observations on a particular day and use them for load flow calculations. Fig.4(a) shows the average load demand data set for a particular week of the year. The figure shows that the load demand was higher for weekdays than in the weekends.Fig.4(b) represents the average load demand data for each month derived from the 15-minute intraday data set. The figure shows that the load demand is less for the summer months when compared to the winter months. The results of the power flow for both the test cases have been shown below in the following section.

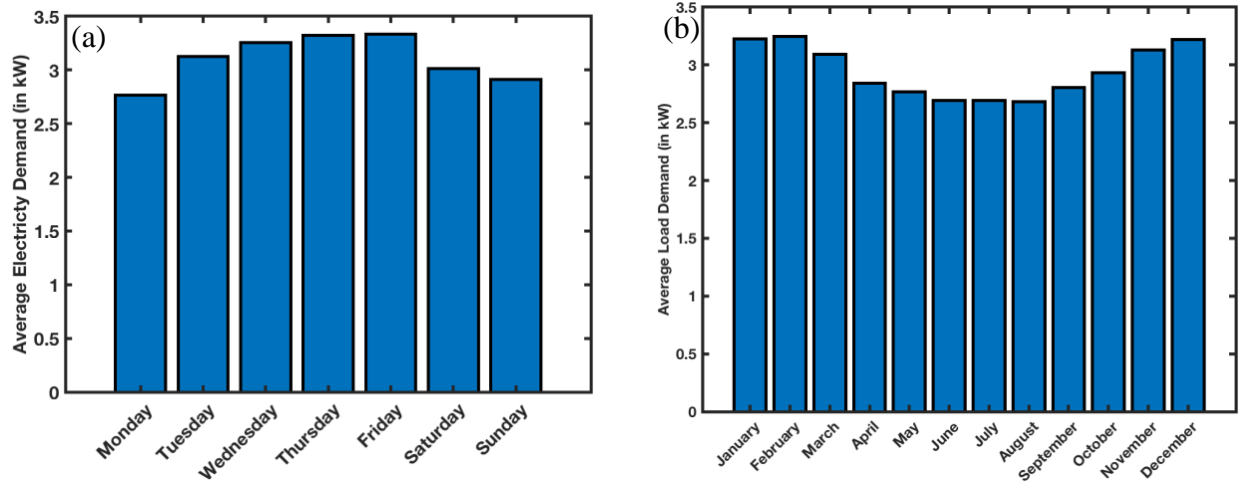

Fig.4. Load Demand Dataset for (a). Day of a week and (b) Average Monthly Demand

## 6.Results and Discussion

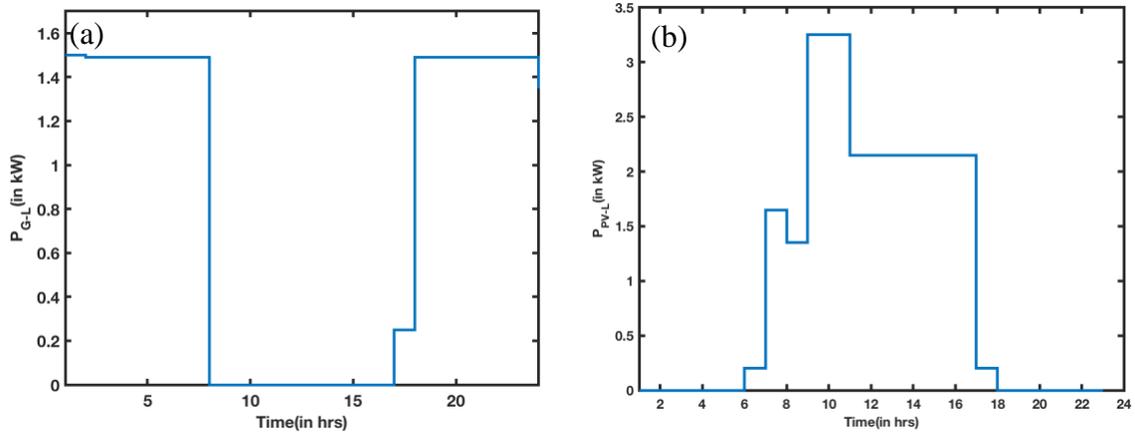

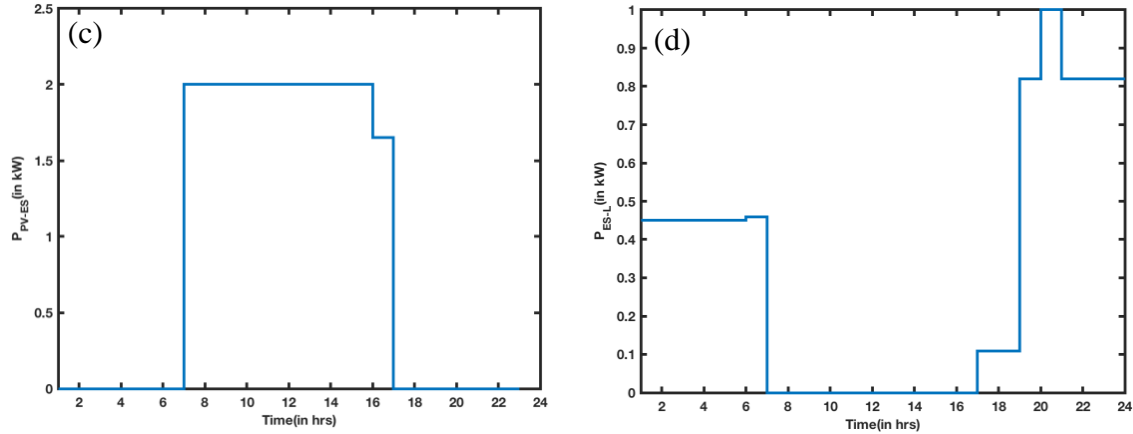

Fig.5. Power Flow Profile for (a). Generator to Load,(b) PV to Load, (c) PV to Battery and (d) Battery to Load for a summer weekday

Fig 5. (a-d) shows the energy flow guided by the optimization process for a 24-h period during a summer day. The overall load demand is met by the DG, PV and the battery during the early hours of the day and the late hours of the evening. According to Fig.5(a), the load demand is met by the DG during the early and the late hours of the day when the battery is under discharge mode. The DG is not functional during the noon when it switches off and the load demand is met by the PV. When the load demand falls below the PV output, the excess energy flows from the PV to charge the battery. Fig.5(b) shows that the operational characteristics of the PV during the mid-day hours especially when the irradiation of the sun is maximum. The PV is non-functional during the night in absence of solar irradiation when the load demand is met by the DG and the battery. Fig.5(c) shows that the energy flow from the PV to the battery. The battery bank is charged during the day which is supplied to the load during the night when in PV is absent. Fig. 5(d) shows the energy flow from the battery to the load on a summer weekday. The battery is supported by the DG in the early hours of the morning and by the PV in the late hours of the night, for meeting the load demand. The scheme described in Eq. 21 switches the state of the battery system from charging to discharging mode when the SOC of the battery is below the minimum limit and vice versa. Thus, it ensures complete absence of simultaneous charging and discharging components in the whole system during the simulation.

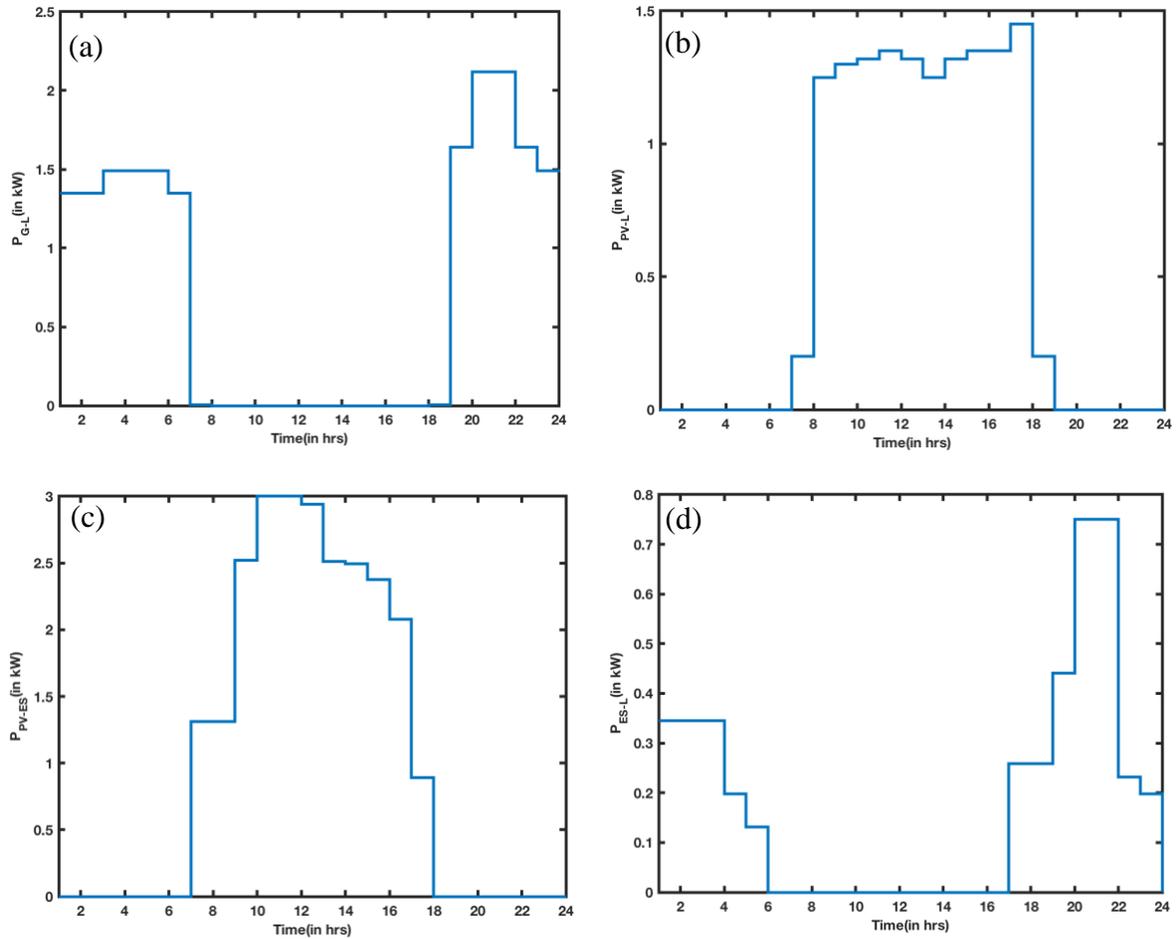

Fig.6. Power Flow Profile for (a). Generator to Load, (b) PV to Load, (c) PV to Battery and (d) Battery to Load for a winter weekday

Fig 6.(a-d) shows the energy flow guided by the optimization process for a 24-h period during a winter weekday. For the winter, the trend is similar to that of the case in summer. Fig. 6(a) shows the energy flow from the DG to the load. It shows that the DG starts earlier and switches off at a later instant when compared to the power flow profile in the previous case. The power flow from the DG to the load is higher in the night than during the day when compared to that of the summer months. Fig.6(b) shows that since the winter months have low irradiation throughout the day, the power flow from the PV to the load are lower than that of the summer (Fig.3(b)). Fig.6(c) shows the power flow from the PV to the battery. During early hours of the day, since the load demand is met by the DG, the power flow from the PV to the battery is lower than that of the summer months. According to Fig.3(b), since the load demand during the day is higher when compared to the summer, the peak energy available for charging the battery is relatively higher. Fig.6(d)

shows the energy flow from the battery to the load. Since the load demand at night is more in the winter season than in the summer, the battery discharges more during the late hours of the night in the summer than in winter. Hence the power flow guided by the cost model accurately portrays the contribution of the energy storage device and the PV under intermittent renewable resources towards the operation of the hybrid system.

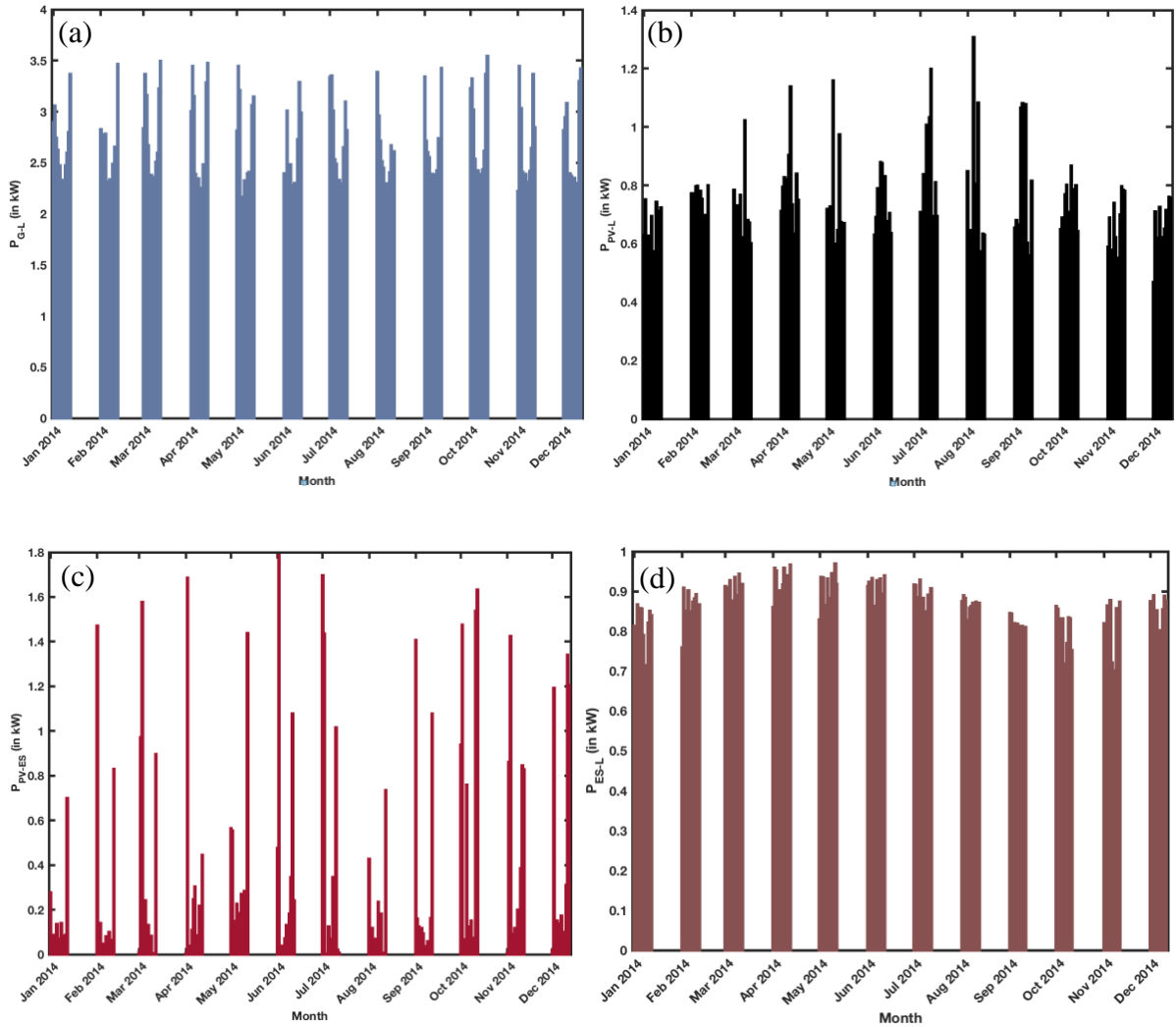

Fig.7. Power Flow Profile for (a). Generator to Load,v(b) PV to Load, (c) PV to Battery and (d) Battery to Load for the whole year

Fig 7. (a-d) shows the energy flow guided by the optimization process for the second data set. Fig. 7(a) shows that the energy flow from the DG to the load decreases during the summer months when compared to that of the winter months. middle. Since the load demand is higher for winter (Fig.4(b)) than in summer,

the energy flow from DG to the load is higher for the winter months when compared to the summer. Since the load demand is higher for weekdays (Fig.4(a)) than in weekends, the energy flow from the DG to the load is higher for the weekdays than during the weekends. Since majority of the months for the year under consideration, have weekdays as their starting and ending days, the energy flow from the DG to the load is higher towards the extremes when compared to the middle of each month. The power flow from the PV to the load for different months of the year has been plotted in Fig.7(b). Since the solar irradiation is higher for the summer seasons than in the winter, the power flow from the PV to the load is higher for the summer months when compared to that in winter. Fig.7(c) shows the power flow from the PV to the battery for each month. The figure further shows that both the power flow at the beginning and the end of each months follows a pattern of seasonality on a quarterly basis. Fig.7(d) also shows a pattern of seasonality on an annual basis. Due to high load demand during the summer and the winter seasons (Fig.4(b)), the figure further shows that the battery gets discharged more during the summer and the early winter season when compared to the rest of the months. From the results, it can be inferred that the model captures both quarterly and annual seasonality existing in the power flow profiles and they correlate with the variation of load demand throughout the year. Thus the proposed formulation captures the underlying characteristics existing in large electrical energy consumption datasets.

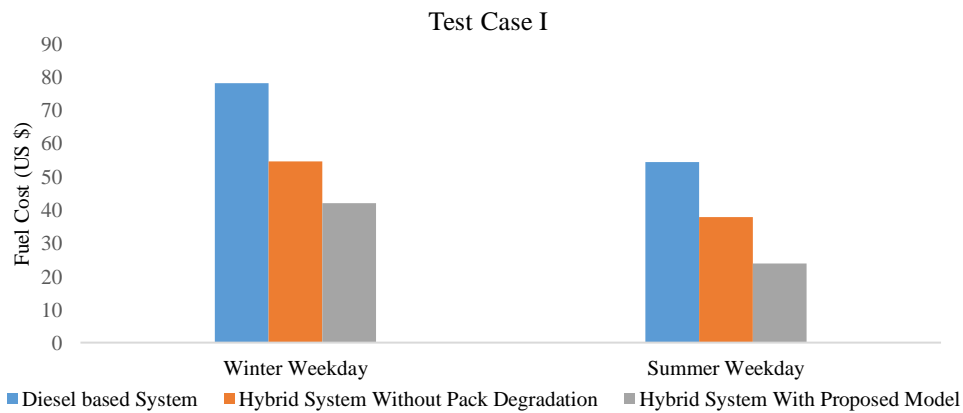

Figure 8. Fuel Cost Savings (in US $) for three different scenarios for Test Case I

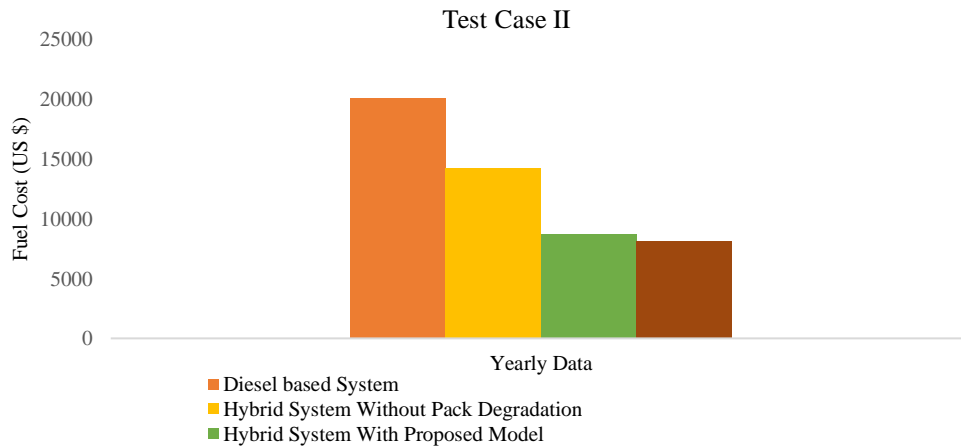

Figure 9. Fuel Cost Savings (in US $) for different scenarios for Test Case II

Fig.8. shows the variation of the total cost of the objective for the first data set under three different cases under consideration namely:- 1. A generic diesel generator based system, 2. A hybrid system where the energy storage components have no degradation cost included and 3. The proposed model with dynamic degradation costs. Compared to the minimum cost obtained for the hybrid system without degradation, the fuel cost reduced by 24.11% and 37.04% for winter and summer respectively and coompared to the minimum cost obtained for the generic DG based system, the fuel cost reduced by 46.97% and 56.09% for winter and summer respectively. Fig.9. shows a similar cost reduction for the three cases for the larger data set. The plot further includes the solution retrieved from the commercially available BARON$^{TM}$ Optimization software [49] which provides algorithms for achieving globally optimal solutions to constrained non-linear optimization problems. For the larger data set, compared to the hybrid model without degradation and the DG based system, the net cost of the objective was reduced by 56.7% and 39.04% for the whole year respectively. The result further shows that there exists a deviation of 6.87% between the global solution by the commercial package and the solution achieved by the algorithm. It can thus be inferred that capacity fade affects the cost solution of the hybrid system and the proposed cost model achieves a lower cost when compared to the presented models of economic dispatch.

## 6. Limitations and Scope for Future Research

The proposed model in the paper does not consider effects of variation of temperature on the charging and discharge efficiencies of the battery. Since the model considers capturing seasonal variations, frequent temperature fluctuations can affect the load demand. Future research will focus on studying how temperature variations affect these semi-empirical models of available capacity of the storage unit. We will be focusing more on capturing these fluctuations by designing convex degradation maps between the temperature fluctuations and charge/discharge efficiencies of the system. A naïve ADMM approach has been chosen to solve the proposed model which offers more accuracy but introduces latency in calculations due to a large number of iterations. Since the model focuses on introducing the convexity of charge/discharge cost functions, we have ignored the run-times of the algorithm. The overall non-linear model is convex and has been solved in a distributed manner. The future research will focus on designing faster distributed techniques for improving upon the convergence rates of the system.

## 7. Conclusion

This paper introduces an optimal pathway towards solving a non-linear least cost minimization problem which incorporates storage dynamics and intermittent renewable generation. The non-linear model incorporates a non-linear degradation cost function which is expressed as an empirical relation between power flow and capacity, thereby making it feasible for plug and play operations during cost analysis in power system applications. The paper proves the convexity of the related models which holds till the Ragone parameter ($\alpha P$) increases to the order of $10^{-6}$. The non-linear model was tested on a load demand data set which spans over a period of 24 hours for two different seasons. It was concluded that compared to the generic model without incorporating cost of degradation and intermittent renewable generation, the fuel cost savings increased drastically for both the test cases under consideration. The model thus provides a more accurate estimate of daily cost savings when compared to the generic stochastic techniques of economy dispatch under cell degradation. Further research will mainly focus on the design of distributed optimization algorithms for solving such non-linear optimization problems with low latency and higher accuracy.

## Acknowledgements

The authors would like to thank Prof. Dr.Yinliang Xu from Tsinghua Berkeley Shenzen Institute for providing valuable feedback and guidance behind the formulation. The authors would also like to acknowledge Dr. Henerica Tazvinga's research in designing models based on minimum cost solutions for energy sectors. Their extensive solar irradiation data set has been extremely useful in the validation of the proposed model.